\documentclass[aps,pre,twocolumn,superscriptaddress]{revtex4-1}
\usepackage[dvips]{graphicx}
\usepackage{afterpage}
\begin{document}

\title{Pedestrian flow through multiple bottlenecks}


\author{Takahiro Ezaki}
\email{ezaki@jamology.rcast.u-tokyo.ac.jp}
\affiliation{Department of Aeronautics and Astronautics, School of Engineering,
The University of Tokyo, 7-3-1, Hongo, Bunkyo-ku, Tokyo 113-8656, Japan}

\author{Daichi Yanagisawa}
\email{daichi@mx.ibaraki.ac.jp}
\affiliation{College of Science, Ibaraki University,
 2-1-1, Bunkyo, Mito, Ibaraki, 310-8512, Japan}

\author{Katsuhiro Nishinari}
\email{tknishi@mail.ecc.u-tokyo.ac.jp}
\affiliation{Research Center for Advanced Science and Technology,
The University of Tokyo, 4-6-1, Komaba, Meguro-ku, Tokyo 153-8904, Japan}


\date{\today}

\begin{abstract}
 We investigate the dynamics of the evacuation process with multiple bottlenecks using the floor field model.
 To deal with this problem, we first focus on a part of the system and report its microscopic behavior.
The system is controlled by parameters of inflow and the competitiveness of the pedestrians,
and large inflow leads to a congested situation. 
Through simulations, the metastable state induced by conflicts of pedestrians is observed.
The metastability is related to the phase transition from free flow to congestion.
The critical condition of the transition is theoretically derived.

In addition, we give simulation results of situations with multiple bottlenecks. They imply that local improvement of 
pedestrian flow sometimes adversely affects the total evacuation time, and that the total optimization of the system is 
not straightforward.
\end{abstract}


\maketitle

\section{Introduction}
\begin{figure}[htbp]
 \begin{center}
  \includegraphics[width=70mm]{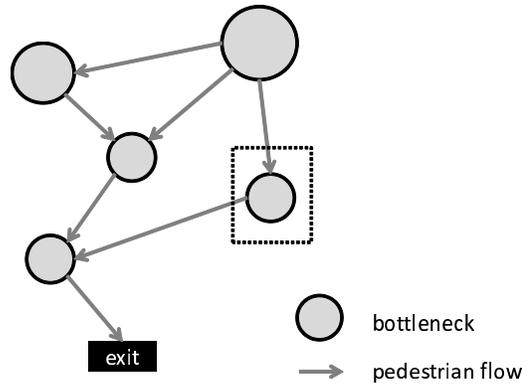}
 \end{center}
 \caption{Schematic diagram of pedestrian evacuation from a building. A single segment enclosed with dotted lines is mainly focused on.}
 \label{concept}
\end{figure}

Dynamical behavior of crowds has attracted many physicists over the last decades for its 
nontrivial characteristics \cite{ped,ped2}. The motion of pedestrians can be regarded as a problem of a many-body system of 
``self-driven" particles. In order to investigate the collective phenomena of the system, many microscopic models have been developed: 
the social force model \cite{SF}, the floor field (FF) model \cite{FF,FFc,FF2,FFy,FFy2,FFy3,PFF,AFF,FFF}, the lattice gas model \cite{LG}, etc. In addition to the simulations, much effort has also been 
devoted to experimental studies \cite{ex,ex2}.
In this research field, the evacuation of crowds has been vigorously studied since it is of great importance to design 
buildings properly for the case of emergency, in the context of risk management \cite{man}. One remarkable phenomenon
observed during evacuation is the clustering of pedestrians at bottlenecks such as exits (\textit{arching}). When more than one pedestrian tries to move to the same place,
the \textit{conflict} occurs, decreasing the outflow of pedestrians. 
The effect of the conflict is a dominant factor of total evacuation time; however, only a few theoretical analyses have been performed so far \cite{FF2,FFy,FFy2,FFy3}.
In previous studies, the evacuation process from a single room has mainly been investigated. However, in an actual emergency, the pedestrian flow
 experiences many bottlenecks and merges together toward the exit. To the authors' knowledge, no systematic approach to evacuation from complex buildings
has been proposed. This study provides a first step toward the understanding of the problem. First, as illustrated in Fig. \ref{concept}, we abstract the most important factors, i.e., bottlenecks
 and their connectivity as a network. By considering this kind of a general network, one can apply our results to a broad range of practical problems.
 In this study, different from most other studies of networks, each node itself has a complex dynamics; we therefore begin by focusing on a single segment of the network.

In this paper, the overall argument is based on the FF model, which is one of the well-established cellular automata models for describing 
the pedestrian dynamics. The model is convenient not only for simplicity and ease of use, but for its extensibility \cite{FFc,FF2,FFy,FFy2,FFy3,PFF,AFF,FFF}.

The effect of the conflict was first implemented in the FF model by introducing the friction parameter in Refs. \cite{FFc,FF2}.
In the present study, we use a more general version of the friction parameter, namely, the \textit{friction function} which was first proposed in Ref. \cite{FFy2}.
The motion of pedestrians involved in the conflict is canceled with a certain probability determined by the friction function, 
which controls the strength of clogging and sticking among pedestrians. 

In addition to exits, we set an entrance providing the system with pedestrians with a certain probability every time step. 
The stochastic entrance is convenient for us to control the inflow of pedestrians \cite{FFy3,PFF}, and by regarding it as inflow from an exit of the previous bottleneck,
we can evaluate the effect of connectivity of the bottlenecks.

The rest of this paper is organized as follows. Section \ref{mod} gives the definition of the model. In Sec. \ref{sim}, simulation results are shown.
To explain the phenomenon analytically, we propose the second-order cluster approximation in Sec. \ref{theo}. Finally, we summarize the argument in Sec. \ref{con}.

\section{Model}\label{mod}
\subsection{Floor field model}
\begin{figure}[htbp]
 \begin{center}
  \includegraphics[width=70mm]{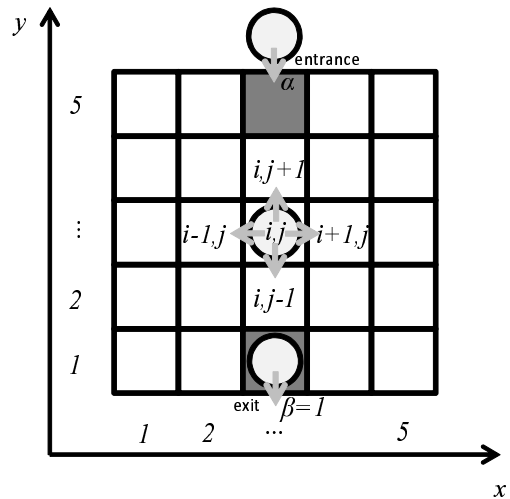}
 \end{center}
 \caption{Update rules. Each pedestrian can hop to its neighboring sites or stay at its present site in a time step. 
 Pedestrians enter the area from the entrance with the probability $\alpha$, and leave the area from the exit with the probability $\beta=1$.}
 \label{jump}
\end{figure}

We consider a two-dimensional lattice representing a room with an entrance and exit, consisting of $N \times N$ sites labeled ($i,j$) ($i,j=1,2,\cdots,N$).
Each site can contain only one pedestrian at most. 
Every time step pedestrians choose one destination site out of their five neighboring sites including the present site
: ($i,j$), ($i+1,j$), ($i-1,j$), ($i,j+1$), and ($i,j-1$) (see Fig. \ref{jump}),
according to two types of FFs. One of the FFs is the static FF ($S_{ij}$) describing the shortest distance to the exit site, and the other is the dynamic FF ($D_{ij}$) expressing 
the total number of pedestrians who have visited the site. The dynamic FF has the dynamics of diffusion and decay, unlike the static FF \cite{FF}.
The transition probability $p_{ij}$ for a jump to the neighboring site ($i,j$) is determined by the following expression:
\begin{equation}
p_{ij} = Z \xi_{ij}\exp{(-k_sS_{ij}+k_dD_{ij})},
\end{equation}
where $k_s$ and $k_d$ are non-negative sensitivity parameters, and $Z$ stands for the normalization factor. $\xi$ returns $0$ for forbidden transitions such as to a wall, 
an obstacle, and neighboring occupied sites, and returns $1$ for other transitions. 
In this paper, the static FF is given by the $L^2$ norm as
\begin{equation}
S_{ij} = \sqrt{(x_{ij}-x_{\rm{ex}})^2+(y_{ij}-y_{\rm{ex}})^2},
\end{equation}
where ($x_{ij},y_{ij}$) and ($x_{\rm{ex}},y_{\rm{ex}}$) are the coordinates of the site ($i,j$) and the exit site, respectively.
On the other hand, we ignore the effect of the dynamic FF ($k_d=0$) since it does not greatly affect the arguments in this work \cite{ig}.
\subsection{Conflict resolution}%
Due to the use of parallel update it happens that more than one pedestrian tries to choose the same site, which is called the conflict.
The simplest solution of the conflict is to choose one pedestrian randomly to move to the site, and keep other pedestrians at their present sites.
However, in actual situations, pedestrian flow is often clogged by more than one pedestrian moving at the same time.
To model this effect, the friction parameter has been introduced \cite{FFc,FF2},
and many significant results have been obtained so far.
In a recent study \cite{FFy2}, the friction function has been proposed to describe the effect more precisely.
In the friction function, the number of pedestrians involved in the conflict, $k$, is reflected to its resolution probability. In this paper, we assume $\phi$ in
the following form as in Ref. \cite{FFy2,FFy3}:
\begin{equation}
\phi(\zeta,k) = 1 - (1-\zeta)^k-k\zeta(1-\zeta)^{k-1}.
\end{equation}
Here, $\zeta \in [0,1]$ is the friction coefficient representing the strength of the clogging irrelevant to $k$.
This $\phi$ is a monotonically increasing function of $k$ and $\zeta$. 
Note that this choice of $\phi$ is one of the possible expressions. 
If one takes the friction function independent of $k$, it coincides with the friction parameter.

Each conflict is resolved with probability $1-\phi(\zeta,k)$, and one of $k$ pedestrians is randomly selected to move to the site,
otherwise, the conflict remains.

\subsection{Entrance and exit}
In each time step, a pedestrian is provided to an entrance site with the probability $\alpha \in [0,1]$ if the site is empty, and
removed from an exit site with probability $\beta$. (See Fig. \ref{jump}.) In this paper, we assume that each room has enough space so that
pedestrians at exit sites are smoothly accepted to the next room, namely, $\beta=1$.

\begin{figure*}[tbp]
 \begin{center}
  \includegraphics[width=160mm]{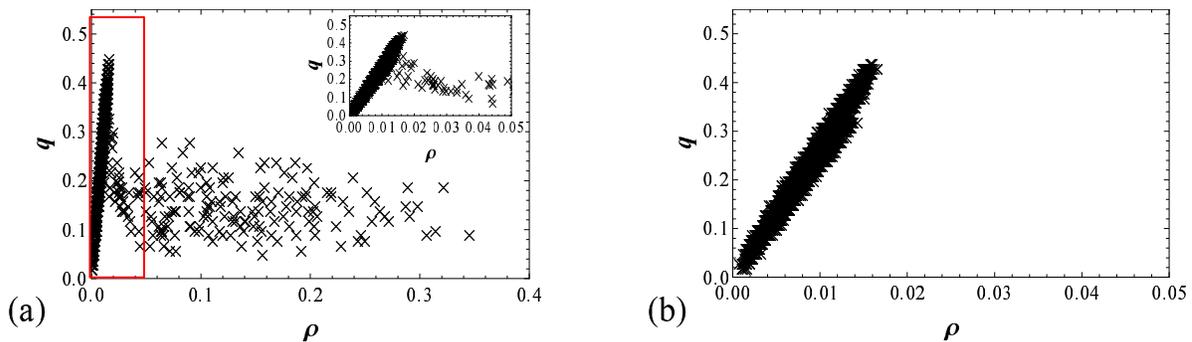}
 \end{center}
 \caption{(Color online.) Fundamental diagram of the system in the competitive parameter regime [(a)$\zeta=0.8$] and the cooperative parameter regime [(b)$\zeta=0.0$].
 Each plot is obtained by setting the inflow probability $\alpha=0.1,0.2,\cdots,0.6$ and averaging the flux and density over 100 time steps.}
 \label{FD}
\end{figure*}

\begin{figure}[tbp]
 \begin{center}
  \includegraphics[width=70mm]{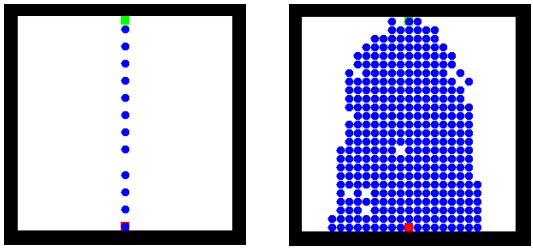}
 \end{center}
 \caption{(Color online.) Snapshots of the simulations. A green site (at the top) and a red site (at the bottom) indicate the entrance and exit sites, respectively. With the same inflow rate, one can
 observe the free-flow phase (left) and congestion phase (right) of pedestrians represented by blue circles.}
 \label{sch}
\end{figure}
\section{Simulations}\label{sim}
In the following, we set $k_s=10$ \cite{ks}.
The dimensions of the simulation area are $25\times 25$, and one entrance site and one exit site are set at $(13,25)$ and $(13,1)$, respectively.
In this section we see some simulation results, varying the inflow probability $\alpha$ and the conflict coefficient $\zeta$.

\subsection{Metastability of pedestrian flow}
In Fig. \ref{FD}, the average pedestrian outflow through the exit, $q$, is plotted. Here, the density $\rho$ is defined as the number of pedestrians in the room divided by
the number of sites in the area, $25 \times 25$. We performed simulations for 100000 time steps for each inflow probability $\alpha = 0.1, 0.2, \cdots, 0.6$ with the initial condition that no pedestrian is set in the system.
The simulation results of the flux and density are averaged over every $100$ time steps. 
The relation between the flux and density is often referred to as
the \textit{fundamental diagram} in the context of traffic flow, and in vehicular traffic the \textit{metastable state} is observed in the 
fundamental diagram. The metastable state indicates an unstable state with high flux and density before falling to the jammed state, and at the same density we can see 
multiple fluxes corresponding to the metastable state and jammed state, in a certain regime of density.
Interestingly, the metastability is observed also in this problem. 
While in vehicular traffic flow, the metastability comes from the effect of inertia of vehicles; in this case, the conflict plays an important role for pedestrian traffic.
If the conflict occurs at the exit, outflow is decreased; therefore, even for the small density of pedestrians in the system, 
the jammed flux can be observed by the concentration of pedestrians at the exit. 
On the other hand, if the pedestrians are dispersed and in the state of free flow, the system can keep large flux. 
This is the mechanism by which the metastability is induced in the pedestrian bottleneck .
Consistent with these arguments, the metastability cannot be observed when $\zeta=0$ [Fig. \ref{FD}(b)].

Next, let us explain the fundamental behavior of the system.
First, the free-flow phase (see Fig. \ref{sch}) can be observed for every $\alpha$ and $\zeta$. 
In this phase, large $\alpha$ directly leads to large flux, corresponding to the linear relation between $q$ and $\rho$.
On the other hand, when the conflict occurs at the exit, the rate of outflow shrinks. 
If this outflow is smaller than inflow (for large $\alpha$ and $\zeta$), the density of pedestrians increases leading to congestion [the congestion phase shown in Fig. \ref{sch} (right)].
In contrast, if the inflow is not enough large, the conflict disappears, recovering the free flow.

Here, if we take $\alpha$ as a controllable parameter for a given $\zeta$, how can we optimize the inflow parameter? 
The answer is to keep the inflow lower than the critical rate, which prevents the conflict at the exit from growing. 
We discuss this issue in the following sections in detail.

\begin{figure*}[htbp]
 \begin{center}
  \includegraphics[width=160mm]{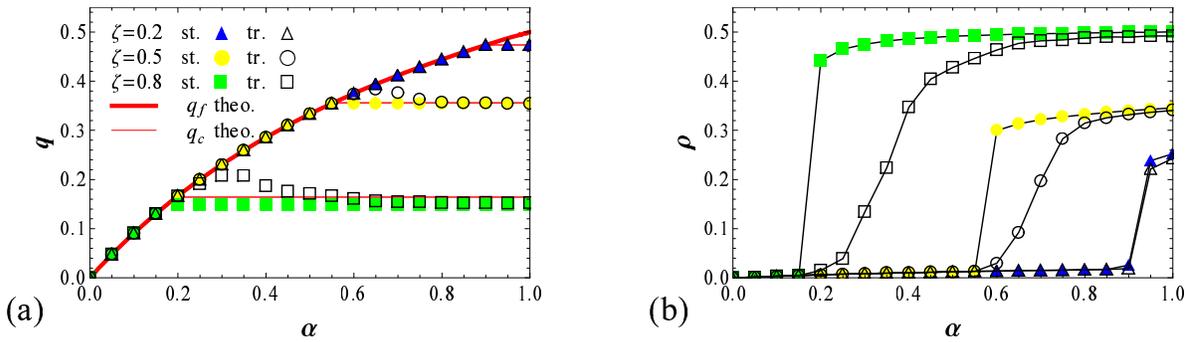}
 \end{center}
 \caption{(Color online.) Pedestrian flux and density in the steady state (st.) and transient state (tr.) vs $\alpha$.  
 Equations (5) and (12) derived in Sec. \ref{theo} are also shown as $q_f$ theo. and $q_c$ theo., respectively.
 The curves are added to improve visibility in  (b).
 }
 \label{RF}
\end{figure*}
\begin{figure}[htbp]
 \begin{center}
  \includegraphics[width=70mm]{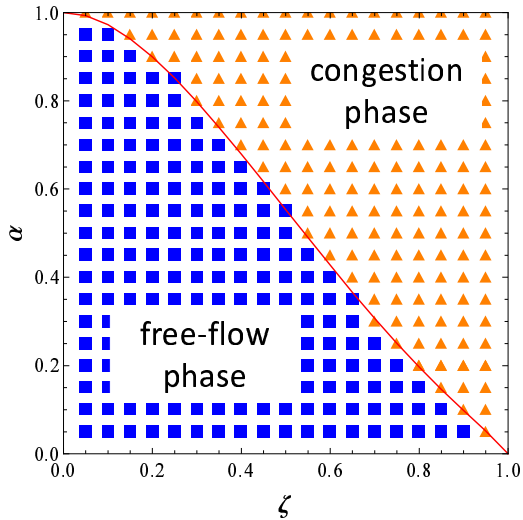}
 \end{center}
 \caption{(Color online.) Phase diagram obtained by simulations (st.). The red line indicates the theoretical critical condition (\ref{a}).}
 \label{acr}
\end{figure}

\subsection{Critical phenomenon related to $\alpha$ and $\zeta$}

In this subsection, we analyze the phase transition from free flow to congestion in detail. 
Figure \ref{RF} shows the density and flux in the steady state and the \textit{transient} state. 
Here, the transient state is defined as the state of the system with free-flow initial conditions (the room with no pedestrian) at finite time steps
 and introduced to include the probability of the system having the phase transition. 
 Even in the parameter regime of the congestion the system might keep the metastable state for finite time steps.
 Hence the expected value of the flux (density) in  the transient state is larger (lower) than that in the steady state.
 These kinds of quantities are also important because the system does not always reach the stationary state in actual evacuation of crowds.
 
 Here, we summarize the simulation conditions in Table \ref{simc}.
 In the simulations, each plot of the transition state (tr.) is obtained by averaging the flux or density over $t_{\rm{max}}=100000$ time steps and $100$ samples 
 \cite{tr}.
On the other hand, plots for the steady state (st.) are calculated by averaging over 1000000 time steps from $t=100000$ to $t_{\rm{max}}=1100000$. For the st. plots, we adopted the 
initial condition that pedestrians occupy all the available sites to ensure the occurrence of the congested situation in the corresponding parameter regime.

In Fig. \ref{RF} (b), the lines of the density jump at each critical $\alpha$, which correspond to the occurrence of the congestion. 
Corresponding to the fact, the flux of pedestrians [Fig. \ref{RF} (a)] agrees with two types of lines: In the free-flow phase, the flux is determined only by the inflow probability;
on the other hand, in the parameter regime of the congestion, the flux is determined by the outflow. These two pedestrian fluxes are evaluated theoretically in the next section.
Moreover, the simulation results of the steady states imply that the situation near the exit can be assumed to be independent of the inflow probability in the congested situation.
On the other hand, one can see the peak of the flux in the transient state. Since the system does not always cause congestion even in the congestion regime of parameters (metastability), the expected flux is higher than the congestion flux for $\alpha$ around the critical value.
If the inflow rate is large enough, the probability of the flux falling to the congested situation increases, and thus, the average flux decreases.

Figure \ref{acr} shows the phase diagram of the system. For each $\zeta$, the upper limit of the inflow rate to keep the free flow is depicted with the line derived in Sec. \ref{cc}.
From a practical standpoint, it corresponds to the criteria for avoiding the congested situation. 

\begin{table}[htb]
  \begin{center}
    \caption{Simulation conditions.}
    \begin{tabular}{c||c|c} \hline\hline
      &st. & tr.  \\ \hline
      averaging time steps&$t=100000-1100000$ & $t=1-100000$ \\
      number of samples&$1$          & $100$   \\
      initial condition&$\rho=1$           & $\rho=0$\\
      \hline\hline
    \end{tabular}
    \label{simc}
  \end{center}
\end{table}

\begin{figure}[h]
 \begin{center}
  \includegraphics[width=70mm]{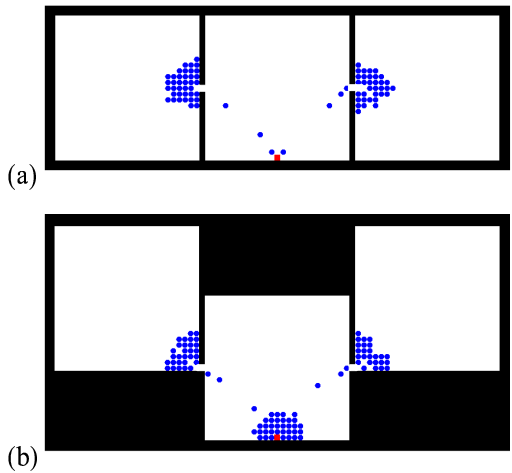}
 \end{center}
 \caption{(Color online.) Snapshots of the merging. (a) Center (ce) connection,  (b) corner (co) connection.}
 \label{ma}
\end{figure}

\subsection{Merging}
\begin{figure*}[tp]
 \begin{center}
  \includegraphics[width=160mm]{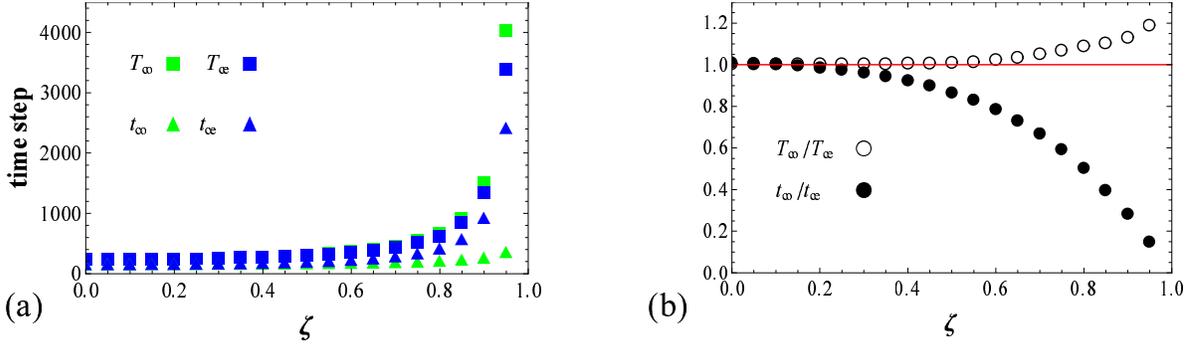}
 \end{center}
 \caption{(Color online.) (a) Average evacuation time and (b) rates of evacuation time. $T$ and $t$ are the average total and local evacuation time, and the subscripts ``co" and ``ce" 
 stand for the conditions of the corner connection and the center connection, respectively. Each plot is calculated by averaging 1000 samples.}
 \label{merging}
\end{figure*}

Buildings have branches of passages, and the merging of the pedestrian flow is an important and complex problem in architectural design.
In this subsection, we present some simulation results to show a paradoxical effect of local improvement of the pedestrian flow.
As more macroscopic systems than those in previous subsections, we consider the systems with three rooms and two types of connections: the center (ce) connection [Fig. \ref{ma}(a)] and the corner (co) connection [Fig. \ref{ma}(b)].
Here, two rooms are connected to a room with an exit site like an entrance hall, and we do not provide pedestrians to each room after setting initial conditions.
In Ref. \cite{FFy}, it has been demonstrated that the exits at the corner has larger capacity of outflow than ones in the center because they have only two 
neighboring sites, and conflicts are reduced.
Therefore, at first sight, the corner connection seems to be better in the sense of swift evacuation. 
In fact, however, it is quite opposite, as shown in Fig. \ref{merging}. 
The simulations have been performed with an initial condition that 50 pedestrians are randomly distributed in each of two rooms other than the entrance hall.
As expected, the evacuation time from each room to the entrance hall is improved by setting the connection at the corner, namely, $t_{co}<t_{ce}$;
however, the total evacuation time worsens as $\zeta$ increases. 
Accordingly, the rate of the local evacuation time $t_{co}/t_{ce}$ decreases while that of the total evacuation time $T_{co}/T_{ce}$ increases.

This phenomenon is explained as follows. The total evacuation time is determined only by the outflow from the exit, and  the local pedestrian flow
to the entrance hall does not influence the total outflow directly. Furthermore, the large inflow to the entrance hall will increase the
density of pedestrians near the exit, which leads to the decrease of the outflow. In other words, if the local inflow is reduced by a strong bottleneck,
the total outflow is improved conversely.

Thus, we can conclude that pedestrian flow should be dispersed not only spatially but also temporally, and the strong bottlenecks might be used for the total optimization.
Note that here we considered the situations that the effective inflow rate to the entrance hall is larger than the exit capacity, namely, they are in the congestion phase as a whole.
If the exit has enough width, the local optimization directly improves the total evacuation time.

\section{Theoretical Analyses}\label{theo}
Let us analyze the pedestrian outflow in each phase and its critical conditions focusing on a single bottleneck.
In this section, we consider the limit $k_s\rightarrow \infty$, where pedestrians surely move to the most desirable site.

\subsection{Free-flow phase}
In the free-flow phase, the pedestrian flux is evaluated by the inflow. 
First, we consider a balance equation at the entrance site:
\begin{equation}
\alpha (1-\rho_{\rm{en}}) = \rho_{\rm{en}}.
\end{equation}
Here, $\rho_{\rm{en}}$ is the probability of finding a pedestrian at the entrance site.
Since we assume $k_s\rightarrow \infty$, a pedestrian at the entrance site surely leaves the site (with probability $1$) in the next time step.
By solving the equation, we can evaluate the flux $q_f$:
\begin{equation}
q_{f} = \alpha(1-\rho_{\rm{en}})=\frac{\alpha}{1+\alpha}.
\end{equation}
This expression is shown in Fig. \ref{RF} (a) and well agrees with the simulation results.

\subsection{Congestion phase}
\begin{figure}[htbp]
 \begin{center}
  \includegraphics[width=50mm]{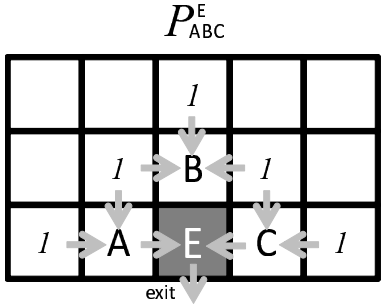}
 \end{center}
 \caption{Cluster approximation. The probability of finding a pedestrian on the second neighboring sites of the exit is assumed to be $1$.}
 \label{appr}
\end{figure}

In the congested situation, the area near the exit is almost fully occupied.
This fact enables us to estimate the outflow in the steady state. 
In the previous studies, the pedestrian density at the neighboring sites of the exit is approximately assumed to be $1$ \cite{FF2,FFy,FFy2}.
However, when the effect of the conflict is strong, this assumption does not give a good estimate.
Especially in this paper, since the effect of the conflict depends on the number of pedestrians involved, we have to take the configurations
of pedestrians at the exit into consideration.
Hence, we adopt a second-order approximation here. In the approximation, we assume the probability of finding a pedestrian at 
the site where the exit is accessible with two jumps (the second neighboring site) to be $1$ (see Fig. \ref{appr}). 
Then, the states of the neighboring sites are characterized by four occupation numbers $A,B,C,$ and $E$, which take $0$ (empty) or $1$ (occupied).
By considering transition probabilities among these states, we can obtain the probability distribution $P^{E}_{ABC}$ of finding each state in
the steady state.
Here, the superscript corresponds to the occupation number of the exit, and the subscripts indicate one of its neighboring sites as shown in Fig. \ref{appr}.
To reduce the dimension of the transition matrix, we use the following facts.
First, we can easily find $P^{0}_{000}=0$ and $P^{1}_{111}=0$. Since no configuration of pedestrians can result in these states in the next time step,
they are not realized in the stationary state.
Furthermore, by symmetry of the system, the equations
\begin{eqnarray}
P^{0}_{100}&=&P^{0}_{001},\\
P^{0}_{110}&=&P^{0}_{011},\\
P^{1}_{100}&=&P^{1}_{001},\\
P^{1}_{110}&=&P^{1}_{011},
\end{eqnarray}
are satisfied. 
With the normalization condition, the transition matrix is summarized as follows:
\begin{widetext}
\begin{equation}
\left(
\begin{array}{c}
P^{0}_{100} \\
P^{0}_{010} \\
P^{0}_{110} \\
P^{0}_{101} \\
P^{0}_{111} \\
P^{1}_{000} \\
P^{1}_{100} \\
P^{1}_{010} \\
P^{1}_{110} \\
P^{1}_{101} \\
\end{array}
\right)
=
\left(
\begin{array}{ccccccccccc}
0 &0 &0 &0 &0 &\frac{1}{4}\phi_2^2 &\frac{1}{2}\phi_2^2 &0 & 0&0&\\
0 & 0& 0& 0& 0&\frac{1}{4}\phi_2^2 &0 &\phi_2^2 &0 &0 &\\
0 &0 & \phi_2^2& 0& 0&\frac{1}{2}\phi_2\tilde{\phi}_2 & \frac{1}{2}\phi_2\tilde{\phi}_2&\phi_2\tilde{\phi}_2 & \phi_2&0 &\\
0 &0 &0 &\phi_2\phi_3 &0 &\frac{1}{4}\phi_3+\frac{1}{2}\phi_2\tilde{\phi}_2 & \phi_3+\phi_2\tilde{\phi_2}&0 &0 &\phi_3 &\\
0 &0 &2\phi_2 \tilde{\phi}_2 & \phi_2\tilde{\phi}_3 &\phi_3 &\frac{3}{4}\tilde{\phi}_2^2+\frac{1}{4}\tilde{\phi}_3 &\tilde{\phi}_2^2+\tilde{\phi}_3 &\tilde{\phi}_2^2 &2\tilde{\phi}_2 &\tilde{\phi}_3 &\\
\phi_2^2 &\phi_2^2&0 &0 &0 &0 &0 &0 & 0 &0&\\
\frac{1}{2}\phi_3+\frac{1}{2}\phi_2\tilde{\phi}_2 & \phi_2\tilde{\phi_2}&\frac{1}{2}\phi_2\tilde{\phi_2}&\frac{1}{2}\tilde{\phi}_2\phi_3 &0 &0 &0 &0 & 0& 0&\\
\phi_2\tilde{\phi}_2 &0 &\phi_2\tilde{\phi}_2 &0 & 0& 0& 0& 0& 0&0 &\\
\frac{1}{2}\tilde{\phi}_2^2+\frac{1}{2}\tilde{\phi}_3 &0 &\frac{1}{2}\tilde{\phi}_2^2&\frac{1}{2}\tilde{\phi}_2\tilde{\phi}_3 & \frac{1}{3}\tilde{\phi}_3 & 0& 0& 0& 0&0 &\\
0 &\tilde{\phi}_2^2 &\tilde{\phi}_2^2 &0 &\frac{1}{3}\tilde{\phi_3} &0 & 0& 0& 0&0 &
 \end{array}
\right)
\left(
\begin{array}{c}
P^{0}_{100} \\
P^{0}_{010} \\
P^{0}_{110} \\
P^{0}_{101} \\
P^{0}_{111} \\
P^{1}_{000} \\
P^{1}_{100} \\
P^{1}_{010} \\
P^{1}_{110} \\
P^{1}_{101} \\
\end{array}
\right)\label{trM}
\end{equation}
\begin{equation}
2P^{0}_{100}+P^{0}_{010}+2P^{0}_{110}+P^{0}_{101}+P^{0}_{111}+P^{1}_{000}+2P^{1}_{100}+P^{1}_{010}+2P^{1}_{110}+P^{1}_{101}=1.\label{nc}
\end{equation}
Note that abbreviated notations $\phi_2 = \phi(\zeta,2),\phi_3 = \phi(\zeta,3),\tilde{\phi}_2=1-\phi_2$, and $\tilde{\phi}_3=1-\phi_3$ are used.
Then, the pedestrian flux is given by
\end{widetext}
\begin{equation}
q_{c} = P^{1}_{000}+2P^{1}_{100}+P^{1}_{010}+2P^{1}_{110}+P^{1}_{101}.
\end{equation}
This expression is illustrated with simulation results in Fig. \ref{RF} (a).
When the effect of the conflict is dominant, the error becomes large. 
In this parameter region, pedestrians are not provided smoothly to the second neighboring sites due to the strong friction,
and thus, the assumption of the approximation is not satisfied.
On the other hand, when the effect of the conflict is not strong, the approximation gives good evaluation.

\subsection{Critical conditions}\label{cc}
In the previous subsections, we evaluated pedestrian flux in the free-flow situation $q_f$ and congested situation $q_c$.
When $q_c$ becomes smaller than $q_f$, the system cannot maintain the free-flow situation; therefore we can obtain the critical condition
regarding the inflow probability for a given $\zeta$ from the condition, $q_c = q_f$:
\begin{equation}
\alpha_{\rm{cr}} = \frac{q_c}{1-q_c}.\label{a}
\end{equation}
This critical condition is compared with simulation results in Fig. \ref{acr}, and well describes the phase transition.

Using this expression, we can control the inflow, preventing the occurrence of the congestion. 
For example, by securing additional routes and keeping the total inflow to one bottleneck lower than the outflow 
in the congested situation, the clustering at the exits can be avoided, which may shorten the total evacuation time.

\section{Conclusions}\label{con}
In this paper, we have taken a first step to treat the problems of multiple bottlenecks in evacuation process.
To consider the problems, we focused on a part of the bottlenecks, using a stochastic entrance.
Through simulations, the metastable state of pedestrian flow arising from the effect of pedestrian conflicts is demonstrated. 
Supported by approximate analyses we have derived the expressions for the pedestrian flux in the free-flow phase and congestion phase, and a critical condition of the inflow
to prevent the congestion. 
Furthermore, interesting phenomena related to the merging of pedestrian flow have been reported. 
The local improvement of the pedestrian flow sometimes causes more serious congestion.
We believe we could give some hints for designing evacuation routes from a theoretical point of view.
To validate these characteristics, some experimental studies would also be necessary.

In this paper, only simple situations have been focused on to concentrate on the essence of the problem.
However, other factors such as the width of exits, multiple entrances and/or exits, route choice of pedestrians, obstacles, etc., should also be 
investigated in combination with our study in future works.

\begin{widetext}
\section*{Appendix: Explicit expression of $q_c$ for a simple case}
Although we can obtain the explicit expression of $q_c$ by solving Eqs. (\ref{trM}) and (\ref{nc}), 
 the expression is  very complicated in general.
If we use the friction parameter $\mu$ \cite{FFc,FF2,FFy} independent of the pedestrian number, namely, $\phi_2=\phi_3=\mu$, 
the expression will be relatively simple. 
Here we show the result in the second-order approximation for this case, $q_2(\mu)$, and compare it with the previous result derived
by first-order approximation \cite{FF2}. From simultaneous Eqs. (\ref{trM}) and (\ref{nc}), we can derive
\begin{equation}
q_2 (\mu) = \frac{48+72\mu-132\mu^2-28\mu^3+140\mu^4-236\mu^5+131\mu^6+49\mu^7-91\mu^8+125\mu^{9}-126\mu^{10}+57\mu^{11}-9\mu^{12}}
{96+192\mu-144\mu^2-68\mu^3+240\mu^4-404\mu^5+78\mu^6+129\mu^7-166\mu^8+185\mu^{9}-117\mu^{10}+48\mu^{11}-9\mu^{12}}.
\end{equation}
On the other hand, Ref. \cite{FF2} presented the expression derived by the first-order approximation, $q_1(\mu)=\frac{1-\mu}{2-\mu}$. The difference
between these equations yields
\begin{eqnarray}
&&q_2(\mu)-q_1(\mu)\nonumber\\
&=& \frac{\mu^5(32+16\mu-84\mu^2+64\mu^3-10\mu^4-75\mu^5+75\mu^6-18\mu^7)}
{(2-\mu)(96+192\mu-144\mu^2-68\mu^3+240\mu^4-404\mu^5+78\mu^6+129\mu^7-166\mu^8+185\mu^{9}-117\mu^{10}+48\mu^{11}-9\mu^{12})}\nonumber\\
&=&\frac{1}{6}\mu^5 - \frac{1}{6}\mu^6+\cdots \qquad (\mu \ll 1)\\.
\end{eqnarray}
Hence $q_2(\mu)=\frac{1-\mu}{2-\mu}+ \frac{1}{6}\mu^5 + O(\mu^6)$ is asymptotic to $q_1(\mu)$ in the $\mu\rightarrow 0$ limit, and we can conclude that
the second-order approximation presented in this paper improves the prediction, especially when the effect of the conflicts is strong.
\end{widetext}
%
%


\begin{thebibliography}{5}
\bibitem{ped}T. Nagatani, Rep. Prog. Phys. 65, 1331 (2002).
\bibitem{ped2}D. Helbing, Rev. Mod. Phys. 73, 1057 (2001).
\bibitem{SF}D. Helbing and P. Molnar, Phys. Rev. E 51, 4282 (1995).
\bibitem{FF}C. Burstedde, K. Klauck, A. Schadschneider, and J. Zittartz, Physica A 295, 507 (2001).
\bibitem{FFc}A. Kirchner, H. Kl\"{u}pfel, A. Schadschneider, K. Nishinari, and M. Schreckenberg, Physica A 324, 689 (2003).
\bibitem{FF2}A. Kirchner, K. Nishinari, and A. Schadschneider, Phys. Rev. E 67, 056122 (2003).
\bibitem{FFy}D. Yanagisawa and K. Nishinari, Phys. Rev. E 76, 061117 (2007).
\bibitem{FFy2}D. Yanagisawa, A. Kimura, A. Tomoeda, R. Nishi, Y. Suma, K. Ohtsuka, and K. Nishinari, Phys. Rev. E 80, 036110 (2009).
\bibitem{FFy3}D. Yanagisawa, R. Nishi, A. Tomoeda, K. Ohtsuka, A. Kimura, Y. Suma, and K. Nishinari, \textit{SICE Journal of Control, Measurement, and System Integration} 3, 395 (2010).
\bibitem{PFF}T. Ezaki, D. Yanagisawa, K. Ohtsuka, and K. Nishinari, Physica A 391, 291 (2012).
\bibitem{AFF}Y. Suma, D. Yanagisawa, and K. Nishinari, Physica A 391, 248 (2012).
\bibitem{FFF}C. M. Henein and T. White, Physica A 373, 694 (2007).
\bibitem{LG}W. Song, X. Xu, B. H. Wang, and S. Ni. Physica A 363, 492 (2006).
\bibitem{ex}T. Kretz, A. Gr\"{u}nebohm, and M. Schreckenberg, J. Stat. Mech. 10014 (2006).
\bibitem{ex2}A. Seyfried, T. Rupprecht, O. Passon, B. Steffen, W. Klingsch, and M. Boltes, Transportation Science 43, 395 (2009).
\bibitem{man}R. W. Perry, M. K. Lindell, and M. R. Greene, \textit{Evacuation Planning in Emergency Management} (Lexington Books, Toronto, 1981).
\bibitem{ig}The arguments are based on the flow rate of pedestrians through entrances and exits, which is not greatly influenced by the dynamic floor field in simple cases. See also \cite{FFy,FFy2,FFy3}.
\bibitem{ks}Here we set ks large enough to make each pedestrian almost surely choose the most desirable site every time step and avoid irrational movement.
On the other hand,
we will approximately assume $k_s\rightarrow \infty$ in theoretical analyses. 
Accordingly, the simulation results with $k_s=10$ and theoretical analyses with $k_s \rightarrow \infty$ are compared in Fig. \ref{RF}.
\bibitem{tr}Strictly speaking, this  quantity corresponds to the `averaged' transient state, rather than the transient state at timestep $t_{max}$.
The quantity is dependent on $t_{max}$ and converges to that of the steady state in the limit of $t_{max}\rightarrow\infty$. 

\end{thebibliography}
\end{document}